\documentclass[fleqn,usenatbib]{mnras}
\usepackage{newtxtext,newtxmath}
\usepackage[T1]{fontenc}
\DeclareRobustCommand{\VAN}[3]{#2}
\let\VANthebibliography\thebibliography
\def\thebibliography{\DeclareRobustCommand{\VAN}[3]{##3}\VANthebibliography}
\usepackage{graphicx}	
\usepackage{amsmath}	
\usepackage{multirow}

\newcommand{\pdm}{\ensuremath{p(\mathrm{DM})}}
\newcommand{\dmcosmic}{\ensuremath{\mathrm{DM_{cosmic}}}}
\newcommand{\cMpc}{\ensuremath{h^{-1}\,\mathrm{Mpc}}}

\newcommand{\zfrb}{\ensuremath{z_\mathrm{frb}}}
\newcommand{\pdmarg}[1]{\ensuremath{p(\mathrm{DM}#1)}}

\newcommand{\thetafb}{\ensuremath{\theta_\mathrm{fb}}}

\title[Neural Network DM Model in CAMELS]{A Neural Network Model for the Cosmic Dispersion Measure in the CAMELS Simulations}

\author[]{Qi Guo,$^{1}$\thanks{E-mail:2463000963@g.ecc.u-tokyo.ac.jp}
 and Khee-Gan Lee$^{1,2}$
\\
$^{1}$Kavli IPMU (WPI), UTIAS, The University of Tokyo, Kashiwa, Chiba 277-8583, Japan\\
$^{2}$Center for Data-Driven Discovery, Kavli IPMU (WPI), UTIAS, The University of Tokyo, Kashiwa, Chiba 277-8583, Japan
}

\date{Accepted XXX. Received YYY; in original form ZZZ}

\pubyear{\the\year{}}

\begin{document}
\label{firstpage}
\pagerange{\pageref{firstpage}--\pageref{lastpage}}
\maketitle

\begin{abstract}
The probability distribution, $p(\mathrm{DM})$ of cosmic dispersion measures (DM) measured in fast radio bursts (FRBs) encodes information about both cosmology and galaxy feedback.
In this work, we study the effect of feedback parameters in the $p(\mathrm{DM})$ calculated from the full Latin Hypercube of parameters sampled by the CAMELS hydrodynamical simulation suite, building a neural network (NN) model that performs well in emulating the effect of feedback on $p(\mathrm{DM})$ at arbitrary redshifts at $z\leq1$.
Using this NN model, we further study the parameter  $F\equiv \sigma_{\rm DM} \, z^{1/2}$, which is commonly used to summarize the scatter on $p(\mathrm{DM})$. We find that $F$ does not depend monotonically on every feedback parameter; instead each feedback mechanism jointly influences the final feedback strength in non-trivial ways.
Even the largest values of $F$ that we find in our entire parameter space are small compared to the current constraints from observed FRB DMs by Baptista et al. 2024, pointing at the limitations of the CAMELS suite due to the small simulation box sizes.
In the future, with larger box-sizes from CAMELS-like suites, similar models can be used to constrain the parameters governing galaxy feedback in the increasing observational samples of FRBs.
\end{abstract}
\begin{keywords}
galaxies: intergalactic medium - transients: fast radio bursts - methods: numerical - software: machine learning.
\end{keywords}

\section{Introduction}
Galaxy feedback processes, such as those from supernovae (SNe) and active galactic nuclei (AGN), have a significant influence on galaxy formation and evolution as well as the distribution of gas in the universe, e.g. across the interstellar medium (ISM), the intergalactic medium (IGM) and the circum-galactic medium (CGM). However, the distribution of gas in different phases, the interaction of hot gas and cold gas, as well as the impact of feedback is complex and our understanding of these processes is incomplete. Cosmological simulations that incorporate baryonic processes, i.e. cosmological hydrodynamical simulations, serve as a crucial tool to study baryonic structures of the observed universe, and as a method to understand the behaviors of the gas. 
For example, early generations of cosmological hydrodynamical simulations were crucial in revealing the warm-hot intergalactic medium (WHIM) phase as the likely reservoir of the so-called missing baryons outside of observed galaxies in the Local Universe \citep{Dave2001,Cen2006}.
Modern cosmological hydrodynamical simulations are constantly improving in terms of box size, resolution, and sophistication of subgrid modeling. The most well-known of these are the so-called `flagship runs', e.g. MillenniumTNG \citep{Pakmor2023}, FLAMINGO \citep{Schaye2023}, SIMBA \citep{Dave2019}, ASTRID \citep{Bird2022}, that usually cover volumes with $\gtrsim 100\,\cMpc$, but
there has also been recent efforts that compare the effect of different feedback models in comparatively smaller boxes. 
\citet{Ayromlou2023} and \citet{Sorini2022}, for example, explored the ejection of baryons from galaxies in the context of various feedback prescriptions within the IllustrisTNG and SIMBA suites.
More recently, \citet{Khrykin2024a} used SIMBA to investigate the effect of stellar and AGN feedback on the global distribution of the cosmic baryons across the IGM and the CGM. 

There are many ways to observationally trace the distribution and properties of the cosmic baryons. For instance, the Lyman-$\alpha$ forest constrains the baryonic matter abundance at $z>2$ \citep{Rauch1997} and can also be applied to perform 3D tomography of the Cosmic Web at these epochs \citep{Lee2016,Lee2018}. 
At lower redshift ($z<1$), however, the growth of WHIM largely upsets the photoionization equilibrium that gives rise to the  Lyman-$\alpha$ forest (although see, e.g., \citealt{Danforth2008}).
Instead, absorption line analysis of various absorbers such as optically-thick HI \citep{Prochaska2017}, CII and CIV \citep{Prochaska2014}, OVI \citep{Simcoe2006}, MgII \citep{Kacprzak2008} and SiIII \citep{Borthakur2016} have been used to place constraints on the baryonic matter distribution residing in various phases, but these are challenging due to the various model assumptions on gas temperature, ionizing background, turbulence, and metallicity that are required to infer a baryon density from the observed tracers.
Moreover, no single absorption line tracer is expected to be able to cover, on its own, a significant fraction of the low-redshift intergalactic gas due to its multi-phase nature.
Given these observational uncertainties, cosmological hydrodynamical simulations have been an important tool for interpretation and making predictions for the properties of the IGM and the CGM. 
For example, traced by OVI, OVII, and OVIII \citep{Nelson2018}, the predictions of the column density distribution functions of the warm–hot phase in IllustrisTNG are consistent with those from observations \citet{Thom2008}. 
Higher resolution cosmological simulations including zoom-in simulations are also run to study the small scale structure of the CGM \citep{Hummels2013,Peeples2019,Hummels2019,Nelson2020,Ramesh2024,Rey2024}.
For instance, \citet{Hummels2013} studied radial profiles of varies of atomic column densities including HI, MgII, SiII, SiIII, SiIV, CIV, NV, OVI and OVII in simulations under the impact of stellar feedback, which could give some constraint on feedback models.

Over the past decade, fast radio bursts (FRBs; see \citealt{Cordes2019}) have emerged as a promising method to trace the properties of cosmic baryons (see review by \citealt{Glowacki:2024}). Manifesting as microsecond to millisecond-duration transient radio pulses, FRB signals undergo dispersion caused by free electrons along the propagation paths. The signal arrival times exhibit a photon frequency dependence ($\Delta t\sim\rm{DM}\nu^{-2}$), where DM is the dispersion measure and represents the integral of electron density over the sightline propagation path:
\begin{equation}
    \mathrm{DM}=\int\frac{n_e}{1+z}\mathrm{d}s,
\end{equation}
where $n_e$ is the physical number density of free electrons and $\mathrm{d}s$ is the proper distance. The total DM signal from an FRB at redshift $z_\mathrm{FRB}$ can be decomposed into several contributions
\begin{equation}
  \mathrm{DM_{FRB}=DM_{MW}+DM_{cosmic}}+\frac{\mathrm{DM_{host}}}{1+z_\mathrm{FRB}},
\end{equation}
where $\rm DM_{MW}$ is the contribution from the Milky Way (MW) ISM and halo gas, $\rm DM_{cosmic}$ is the contribution from diffuse IGM and the intervening galaxy halo CGM, and $\rm DM_{host}$ is the contribution from the host galaxy. 

FRBs can serve as probes of the Large Scale Structure (LSS) and even cosmology.
From the observational side, since the first discovery of FRBs \citep{Lorimer2007}, the number of detected FRBs has increased rapidly thanks to the contributions of different facilities such as CHIME, ASKAP/CRAFT, FAST, DSA, MeerTRAP \citep{CHIME2021,McConnell2016,Nan2011,Kocz2019,Jankowski2022}, and others. By integrating a sample of six localised FRBs, \citet{Macquart2020} showed that the relationship between extra-galactic DMs of these FRBs and redshift is consistent with the baryon density, $\Omega_b$, predicted from the standard $\Lambda \rm CDM$ model albeit with large uncertainty. Using 78 FRBs in which 21 have been localized, \citet{Baptista2024} presented a new measurement of fluctuation parameter $F$ and constraints for the Hubble parameter. From the simulation and theoretical side, \citet{Lee2022} proposed to use FRBs, together with spectroscopic mapping of foreground galaxies, to constrain the baryon fractions and other parameters \citep{Khrykin2024b}. \citet{Batten2021} investigated DM-$z$ relation in Evolution and Assembly of GaLaxies and their Environments (EAGLE) simulations, while \citet{Walker2024} studied the FRB \dmcosmic{} in context of the cosmic web in IllustrisTNG. 

These simulation studies typically analysed the \dmcosmic{} only within a single simulation with a fixed feedback model, but there is increasing interest in investigating the effect on \dmcosmic{} from different feedback models.
\citet{Batten2022} analyzed the \dmcosmic{} distribution, \pdm{} and found that AGN feedback modifies \pdm{} (we drop the subscript `cosmic' for notational brevity when referring to \pdm).
More recently, a recent paper by \citet{Medlock2024} analysed the CAMELS suite, which is 
a large ($>10^3$) suite of small ($L=25\,h^{-1}\mathrm{Mpc}$) cosmological hydrodynamical simulations that sample the parameter space of several cosmological and astrophysical parameters, including the matter density $\Omega_\mathrm{m}$, amplitude of large-scale structure fluctuations ($\sigma_8$), as well as stellar and AGN feedback parameters. 
They analyzed three separate model families covered by CAMELS, i.e. IllustrisTNG, SIMBA, and Astrid in CAMELS and investigated the behaviour of \dmcosmic{} on feedback. They found significant differences in the distribution of \dmcosmic{} between the 3 fiducial models. 
Within each suite, they studied \pdm{} for sightlines up to $z = 1$ in the 1P set under the impact of baryonic feedback.
They found in SIMBA, for the stronger SNe ($A_\mathrm{SN1}=4$) and AGN feedback ($A_\mathrm{AGN1}=4$), \pdm{} is enhanced at lower or higher end of DM, which is contrary to expectation. 
However, \citet{Medlock2024} only carried out 1-dimensional analyses of the 61 ``1P" simulations for each model within CAMELS. that vary one parameter at a time relative to the fiducial model. The full statistical power of the $\sim 1000$ CAMELS realizations have arguably yet to be harnessed to study \dmcosmic{}.

In this paper, we investigate deeper into the CAMELS suite. We computed \dmcosmic{} distribution in SIMBA-CAMELS up to $z=1$. We then train a Neural Network model to compute \dmcosmic{} for arbitrary combinations of model parameters based on the full CAMELS suite of simulations that sample the parameter space as a Latin Hypercube. 
This will be useful in future work to better understand the multi-dimensional effect of feedback processes on \dmcosmic,
and in principle constrain these processes based on the observed  DM-redshift relation. 
Our paper is organized as follows: Section \ref{sec:method} describes the simulation suites, the method to compute DM distribution and the structure of Neural Network. In section \ref{sec:result} we present and briefly discuss the results before concluding in Section \ref{sec:conclusion}.

\section{Method} \label{sec:method}

\subsection{Simulations} \label{sec:simulation}
Cosmology and Astrophysics with Machine Learning Simulations (CAMELS) is a suite of cosmological simulations with a volume of $(25\, h^{-1}\rm {Mpc})^3$ for each box, consisting of 2049 N-body simulations and 2184 hydrodynamic simulations \citep{Villaescusa2021}. 
These hydrodynamic simulations were run with the AREPO and GIZMO codes each with different sub-grid models, following those of the IllustrisTNG \citep{Weinberger2017} and SIMBA \citep{Dave2019} large-volume cosmological simulations. 
These simulations are divided into 4 sets: LH (Latin Hypercube, 1000 simulations), 1P (1-Parameter variations, 61 simulations), CV (Cosmic Variance, 27 simulations) and EX (Extreme, 4 simulations).

The fiducial IllustrisTNG galaxy formation model \citep{Weinberger2017,pillepich2018} includes magneto-hydrodynamics \citep{Pakmor2011,Pakmor2014}. 
It also accounts for the physical processes that simulate galaxy formation and evolution, including the physics of radiative thermochemisty and metal cooling, star formation and evolution, stellar enrichment, supermassive black hole (SMBH) formation, and feedback from stars (supernovae) and SMBH (AGN) along with magnetic fields. 
Meahwhile, the fiducial SIMBA galaxy formation model \citep{Dave2019} accounts for similar processes to IllustrisTNG model but using different models without taking magnetic fields into consideration, while also notably introducing strong AGN jets that are not present in TNG.

Integrated with these two models and designed to train machine learning models, the hydrodynamic simulations span a wide range of cosmological and astrophysical(i.e. feedback) parameters. 
The parameters are $\mathbf{\theta}=(\Omega_\mathrm{m},\sigma_8,A_\mathrm{SN1},A_\mathrm{AGN1},A_\mathrm{SN2},A_\mathrm{AGN2})$, where $\Omega_\mathrm{m}$ is the fraction of matter density, $\sigma_8$ is the r.m.s. of linear matter fluctuations, $A_\mathrm{SN1},A_\mathrm{AGN1}$ and $A_\mathrm{SN2},A_\mathrm{AGN2}$ refer to the strength of SNe feedback and AGN feedback respectively, covering a range of $\Omega_\mathrm{m}\in[0.1,0.5]$, $\sigma_8\in [0.6,1.0]$, $A_\mathrm{SN1},A_\mathrm{AGN1}\in [0.25,4.0]$, $A_\mathrm{SN2},A_\mathrm{AGN2}\in [0.5,2.0]$.
In IllustrisTNG, $A_\mathrm{SN1}$ and $A_\mathrm{SN2}$ control the galactic wind energy emitted per unit of star formation rate and the wind speed respectively, while $A_\mathrm{AGN1}$ represents the energy released in kinetic mode AGN feedback per unit of black hole accretion rate. $A_\mathrm{AGN2}$ represents the ejection speed and burstiness for the kinetic mode of AGN feedback. 
In SIMBA, $A_\mathrm{SN1}$ and $A_\mathrm{SN2}$ control the mass loading factor for stellar feedback and galactic wind speed relative to scalings derived from the FIRE simulations. $A_\mathrm{AGN1}$ parameterizes the momentum flux of kinetic outflows in quasar and jet mode AGN feedback relative to the black hole accretion rate, while $A_\mathrm{AGN2}$ parameterizes the speed of the jet mode AGN feedback. 

More recently, new simulation sets have been released, expanding the ASTRID galaxy formation model and extending the parameter space of the original CAMELS-TNG and CAMELS-SIMBA suites to 28 parameters \citep{Ni2023}. 
However, in this paper we still present the results based on the original CAMELS-TNG and CAMELS-SIMBA suites.

\subsection{Dispersion Measure Distributions}\label{subsec:DM dis}
Our first step is to compute $p(\mathrm{DM})$ in each simulation up to $z=1$ for both CAMELS-SIMBA and CAMELS-TNG. 
We start from \textit{snapshot \#033} ($z=0$) to \textit{snapshot \#018} ($z=1.05$), covering redshift 0.00, 0.05, 0.10, 0.15, 0.21, 0.27, 0.34, 0.40, 0.47, 0.54, 0.61, 0.69, 0.77, 0.86, 0.95 and 1.05.
For each box in CAMELS-SIMBA, we start with the electron density from CAMELS Library\footnote{\url{https://github.com/franciscovillaescusa/CAMELS}}, in which we exclude star formation regions in order to ensure that we are studying only extragalactic baryons. 
Next, similar to the method described in \citet{Batten2021}, we grid each box into $256\times256\times256$ cells and compute the electron density at each grid using pygad\footnote{\url{https://bitbucket.org/broett/pygad/src/master/}} \citep{Rottgers2018}, which bins SPH quantities onto a 3D grid while accounting for smoothing lengths. 
These ensures that the derived properties, including DM distribution are convergent. 
We verified that binning onto different number of grid cells do not change the result appreciably. 
By integrating the electron density of each cell along the direction parallel to a given axis, we produce DM maps from CAMELS-SIMBA.
For CAMELS-TNG, the sightlines are also set to be parallel to one axis of the simulation box.
We use the \texttt{temet} package to ray-trace through the Voronoi cells of the gas distribution (\textcolor{blue}{Nelson et al. in prep}) in the Arepo outputs. 
This yields an ordered list of gas cells intersected by each sightline, and intersection path lengths in a self-consistent way. 
To compute the DM within the box, we again integrate the electron density for each sightline.
Given the ensemble of DM sightlines for a given snapshot in both CAMELS-SIMBA and CAMELS-TNG, we evaluate the probability density functions (PDFs) and cumulative density functions (CDFs) within each snapshot. 

To model $p(\mathrm{DM})$ at redshifts in which the line-of-sight covers a distance larger than one CAMELS box size $(25\, h^{-1} \rm {Mpc})$, we need to connect snapshots together to create continuous lines-of-sight from redshift $z = 0$ to the target redshift $\zfrb$. 
However, the $\Delta z$ between adjacent CAMELS snapshots are far too coarse for us to model line-of-sight redshift evolution by directly connecting adjacent snapshots. 
In order to fill the gaps in between the redshift snapshots, we generate samples covering from $100\, \rm {Mpc}$ to $2400\, h^{-1} \rm {Mpc}$ separated by $L=100\, h^{-1} \rm {Mpc}$ with their redshifts labeled by $z_i$. 
Then we linearly interpolate their CDFs at arbitrary redshifts in between the fixed simulation snapshot redshifts. 
To model the DM of an individual sightline going out to \zfrb{} at a cosmological distance, we draw a random uniform deviate $[0,1]$ at each intermediate redshift $z_i$, then take the corresponding value of the CDF weighted by $z_i$ as the DM value at $z_i$ and sum all the intermediate DMs:
\begin{equation}  
\mathrm{DM}(\zfrb)=\sum_i{n_{e,p}(z_i)}\Delta l_p/{(1+z_i)}=\sum_i{n_{e,c}(z_i)}{(1+z_i)}\Delta l_c,
\end{equation}
where $n_{e,p}=n_{e,c}(1+z)^3$ is the physical electron number density, $\Delta l_p=\Delta l_c(1+z)^{-1}$ is the physical distance.
This method of computing the DM up to a given \zfrb{} by sampling through the CDF is very fast and memory efficient, while implicitly taking into account the sightline diversity to the extent allowable by the CAMELS box sizes.

In order to create the training set for the Neural Network, 
we however need to summarize the \pdm{} for a given CAMELS volume into a succinct input.
We choose to use the fitting function from \citet{Macquart2020}:
\begin{equation}
{p}_{{\rm{c}}{\rm{o}}{\rm{s}}{\rm{m}}{\rm{i}}{\rm{c}}}(\Delta )=A{\Delta }^{-\beta }\exp \,\left[-\frac{{({\Delta }^{-\alpha }-{C}_{0})}^{2}}{2{\alpha }^{2}{\sigma }_{{\rm{D}}{\rm{M}}}^{2}}\right]\,,\,\Delta  > 0
\end{equation}
where $\Delta=\mathrm{DM}/\left \langle \rm{DM} \right \rangle$. 
The nominal free parameters in this function are: $[A,\alpha, \beta, C_0, \sigma_\mathrm{DM}]$.

To infer the fitting parameters for each \pdmarg{|z,\theta}, we run a Markov chain Monte Carlo (MCMC) on the transformed function 
\begin{equation}\label{equ_dm_dis}
f(x)=y^{-\beta} \exp[-\frac{((y^{-\alpha} - C_0)^2)}{2 \alpha^2\sigma_{\rm{DM}}^2}]y,
\end{equation}
where $y = e^{\log(\mathrm{DM})}/\left \langle \rm{DM} \right \rangle$,
to fit the logarithmic distribution of DM and get the best fit parameters $\alpha,\beta$ given $\left \langle \rm{DM} \right \rangle,\sigma_{\rm{DM}}$ which are evaluated directly from each \pdm{} distribution. 
It is worth noting that in this form for the distribution, $A$ and $C_0$ are constrained from the normalizations
\begin{equation}
\int f(\rm{DM})\, \mathrm{d}\mathrm{(DM)}=1,    
\end{equation}
and 
\begin{equation}
  \int \mathrm{DM} f(\rm{DM}) \, \mathrm{d}\mathrm{(DM)}=\left \langle \rm{DM} \right \rangle.  
\end{equation}
Note that specific values or correlations between the fitting parameters, $[A,\alpha, \beta, C_0, \sigma_\mathrm{DM}]$, have been proposed based on physical arguments (see, e.g., \citealt{Macquart2020}.
However, for the purposes of fitting the distributions from CAMELS, we keep full generality for the range allowed for these parameters.

We also explored other methods to compute $p(\mathrm{DM})$ in each snapshot, including the \texttt{yt} package \citep{Turk2011}.
We also tried randomly generating thousands sightlines in each snapshot and then binning the lightrays, followed by calculating the distribution of electron density in each bin and integrating to get the DM. 
After comparing between these methods, we concluded that \texttt{pygad} package works best on SIMBA and \texttt{temet} package works best on TNG.

\subsection{Neural Network (NN)}\label{sec:NN}

In this work, we adopt an approach similar to \citet{Nicola2022}, in which we build the NNs using pytorch\footnote{pytorch:\url{https://pytorch.org/}}, a framework for building deep learning models and optuna\footnote{optuna:\url{https://optuna.org/}}, an automatic hyperparameter optimization software framework. 
We use the LH set of simulations from CAMELS-TNG and CAMELS-SIMBA to train a NN to emulate $p(\mathrm{DM})$ (described by $\left \langle \rm{DM} \right \rangle$, $\sigma_\mathrm{DM}$, $\alpha$ and $\beta$) as a function of cosmological and astrophysical parameters. 
First, we split the simulation boxes from \ref{subsec:DM dis} into training, validation and test sets, which comprise $90\%$, $5\%$ and $5\%$ of the data separately. 
We use all cosmological and astrophysical parameters sampled by the LH, $\mathbf{\theta}=[\Omega_\mathrm{m},\sigma_8,A_\mathrm{SN1},A_\mathrm{AGN1},A_\mathrm{SN2},A_\mathrm{AGN2}]$ as well as redshift $z$ as inputs, while the outputs are the parameters $\left \langle \rm{DM} \right \rangle$, $\sigma_\mathrm{DM}$, $\alpha$, $\beta$ that represent the \pdm. 
Using optuna, we adapt the learning rate, the weight decay, the number of layers, the number of neurons per layer, and the optimizer (Adam, RMSprop and SGD). 
We summarize the structure of the NN below:

\begin{itemize}
    \item Input: $\mathbf{\theta}=(\Omega_\mathrm{m},\sigma_8,A_\mathrm{SN1},A_\mathrm{AGN1},A_\mathrm{SN2},A_\mathrm{AGN2})$ and $z$ 
    \item $N_\mathrm{unit}$ (the number of neurons per layer):[15,35]
    \item $N_\mathrm{lay}$ (Number of layers): [5,15]
    \item Loss function: MSELoss
    \item Activation function: RELU
    \item Output: $\left \langle \rm{DM} \right \rangle$, $\sigma_\mathrm{DM}$, $\alpha$, $\beta$
\end{itemize}

For the hyperparameters, optuna will find a set with the least loss among different runs. 

\section{Results} \label{sec:result}
In a recent study, \citet{Medlock2024} analyzed the behaviors of $p(\mathrm{DM})$ as well as the $F$ parameter in CAMELS-TNG and CAMELS-SIMBA in 1P set.
In this paper we do not reproduce their results but focus on the Neural Network model based on the LH set. 
In this section we first show the result of MCMC fitting for the parametric function of $p(\mathrm{DM})$, followed by a discussion on the performance of the NN.

\subsection{MCMC fitting}\label{mcmc}

\begin{figure*}
    \centering
    \includegraphics[width=1.0\textwidth]{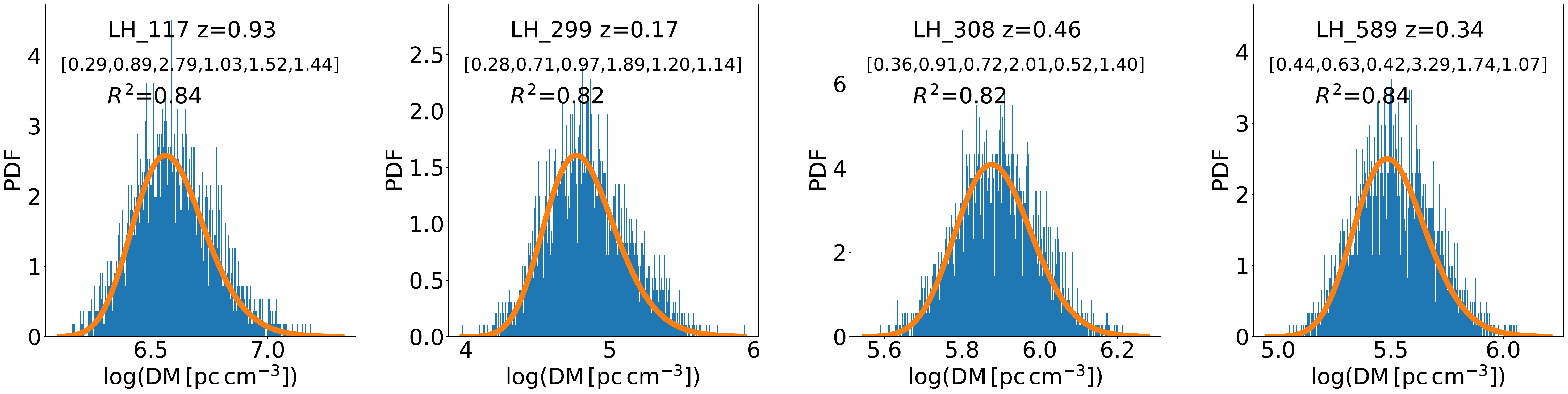}
    \includegraphics[width=1.0\textwidth]{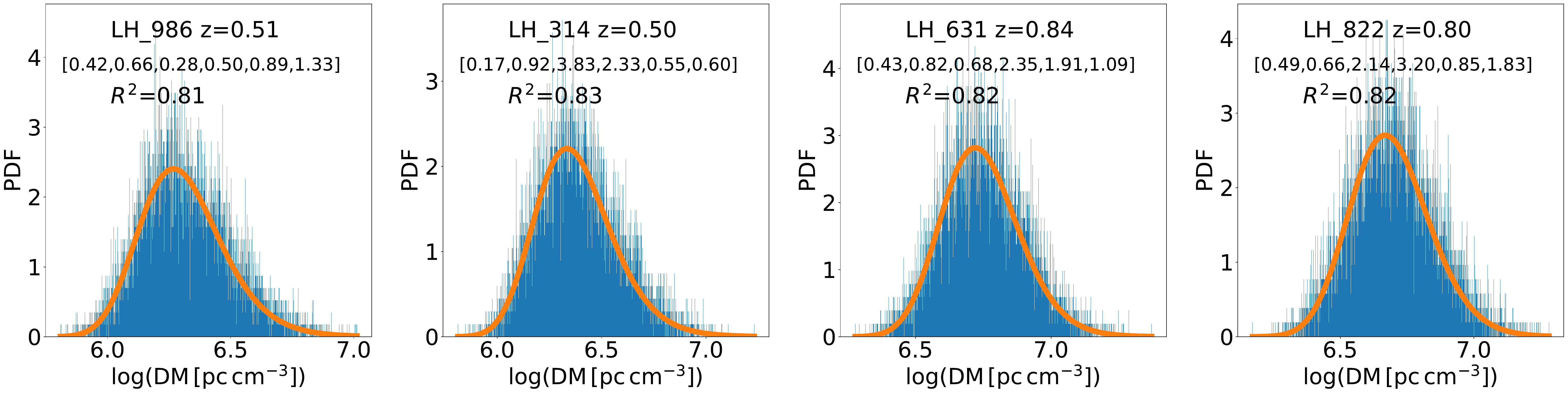}
    \caption{Histograms of $p(\log(\mathrm{DM}))$ (blue) and MCMC fitting results (orange) of several simulations from the LH suite, for SIMBA (upper panels) and TNG (lower panels).
    For these eight panels, LH sets are randomly selected at a random redshift with different cosmological and astrophysical parameters $\mathbf{\theta}=[\Omega_m, \sigma_8, A_{\rm SN1}, A_{\rm AGN1},A_{\rm SN2},A_{\rm AGN2}]$, labeled in the second row of the legend in each panel.
    The orange curves show the fitted results using Equation \ref{equ_dm_dis}, which can describe the \pdm{} fairly well with only a few free parameters.
    }
    \label{fig_SIMBA_MCMC_fitting}
\end{figure*}

Figure \ref{fig_SIMBA_MCMC_fitting} shows the results of MCMC fitting for \pdm{} in SIMBA and TNG suites. 
For simplicity, we randomly select 4 LH sets at a random redshift in TNG and SIMBA with different cosmological and astrophysical parameters. 
The orange curves show the MCMC fitting results based on equation \ref{equ_dm_dis} with $\left \langle \rm{DM} \right \rangle$, $\sigma_\mathrm{DM}$ calculated directly from the distribution and $\alpha$, $\beta$, $C_0$ fitted by \texttt{emcee}. 
We estimate the goodness of fitting based on coefficient of determination defined as
\begin{equation}
    R^2=1-\frac{SS_\mathrm{res}}{SS_\mathrm{tot}},
\end{equation}
where $SS_\mathrm{res}=\sum_i(y_{i}-y_\mathrm{pre,i})^2$ and $SS_\mathrm{tot}=\sum_i(y_{i}-\Bar{y})^2$. 
For nearly all \pdm{}, $R^2\gtrsim0.8$, showing a good performance of the fitting function with respect to the simulated \pdm. 
Using MCMC, with only these few parameters, $p(\mathrm{DM})$ can be well described across the large parameter space and at different redshifts. 
In this work, for simplicity, we use the fitted parameters from MCMC as summary statistics of the overall \pdm{} distributions to train the NN.

\subsection{Performance of NN}\label{sec:NNres}

\begin{figure*}
    \centering
    \includegraphics[width=1.0\textwidth]{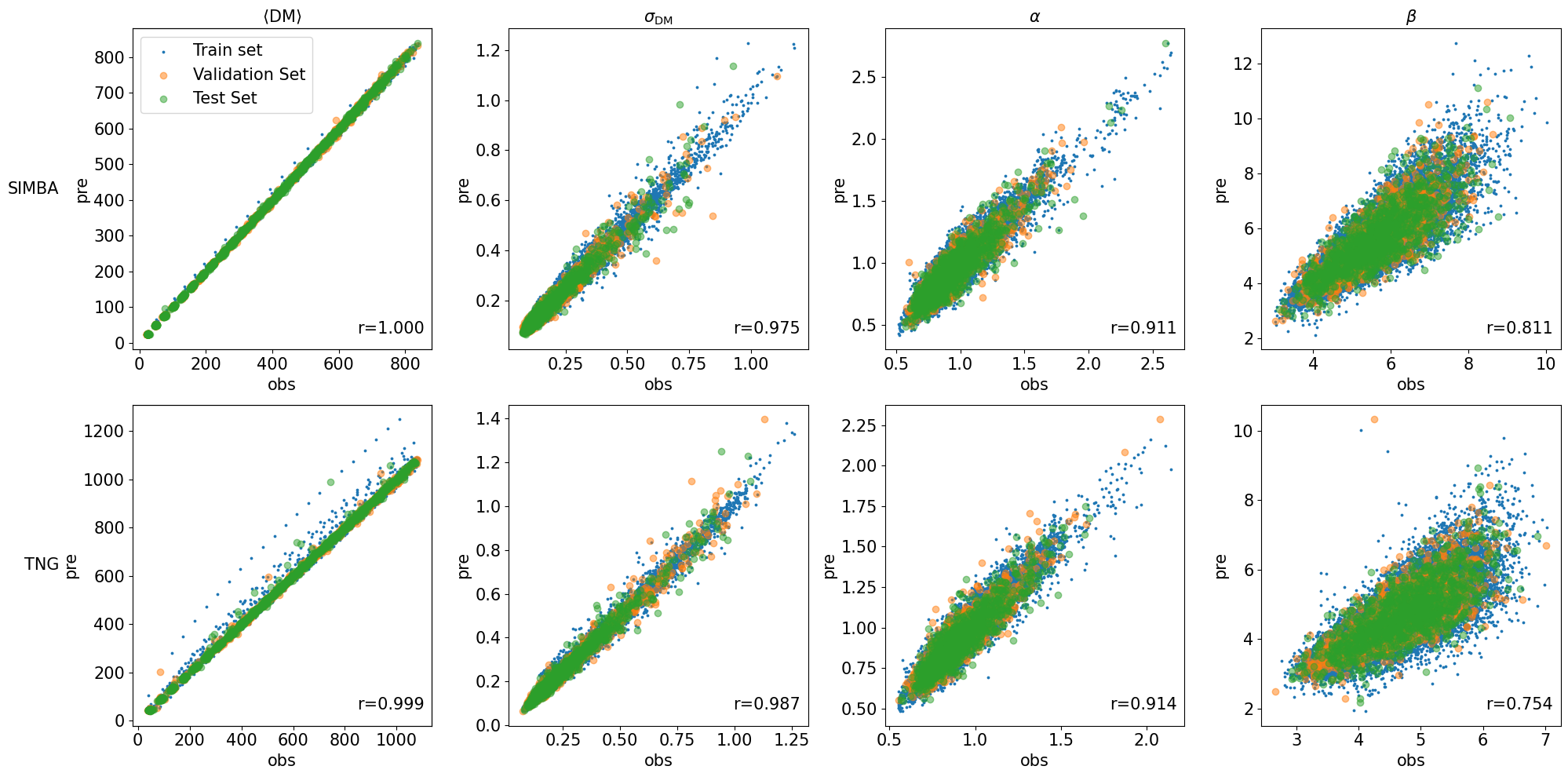}
    \caption{NN predictions of the parameters governing $p(\mathrm{DM})$ for SIMBA (upper panels) and TNG (lower panels). 
    These panels show the true value (obs) fitted from MCMC and predictions (pre) of NN.
    The green circles represent NN predictions on test set, with the Pearson r-coefficients and p values of predicted values and measured results for test sets showing in each panel. 
    Though for $\beta$ the scatter is larger however for $\left \langle \rm{DM} \right \rangle$, $\sigma_{DM}$ and $\alpha$ NN performs well on predicting the results.}
    \label{fig_DM_results}
\end{figure*}

Following the description in §\ref{sec:NN}, we built a NN, mapping from CAMELS parameters $\theta$ and redshift to the fitting parameters describing the functional form for $p(\mathrm{DM})$ (i.e. $\left \langle \rm{DM} \right \rangle$, $\sigma_\mathrm{DM}$, $\alpha$ and $\beta$). 
Figure \ref{fig_DM_results} shows the results of NN for SIMBA and TNG separately, with train set, validation set and test set showing in blue, orange and green circles separately. 
For the test set, we quantify the performance of NN using Pearson correlation coefficients,
\begin{equation}
    r=\frac{\sum_i (y_\mathrm{meas,i}-\Bar{y}_\mathrm{meas,i})(y_\mathrm{pre,i}-\Bar{y}_\mathrm{pre,i})}{\sqrt{\sum_i (y_\mathrm{meas,i}-\Bar{y}_\mathrm{meas,i})^2}\sqrt{(y_\mathrm{pre,i}-\Bar{y}_\mathrm{pre,i})^2}}.
\end{equation}
With high significance, NN makes good predictions on $\left \langle \rm{DM} \right \rangle$, $\sigma_\mathrm{DM}$, $\alpha$ and $\beta$.
However, because of the degeneracy the these parameters and the uncertainties of MCMC fitting on $\alpha$ and $\beta$, the performances on $\beta$ and $\alpha$ are not as good as those on other parameters, with high scatter.

\subsection{$F$ parameter}
\begin{figure*}
    \centering
    \includegraphics[width=0.98\linewidth]{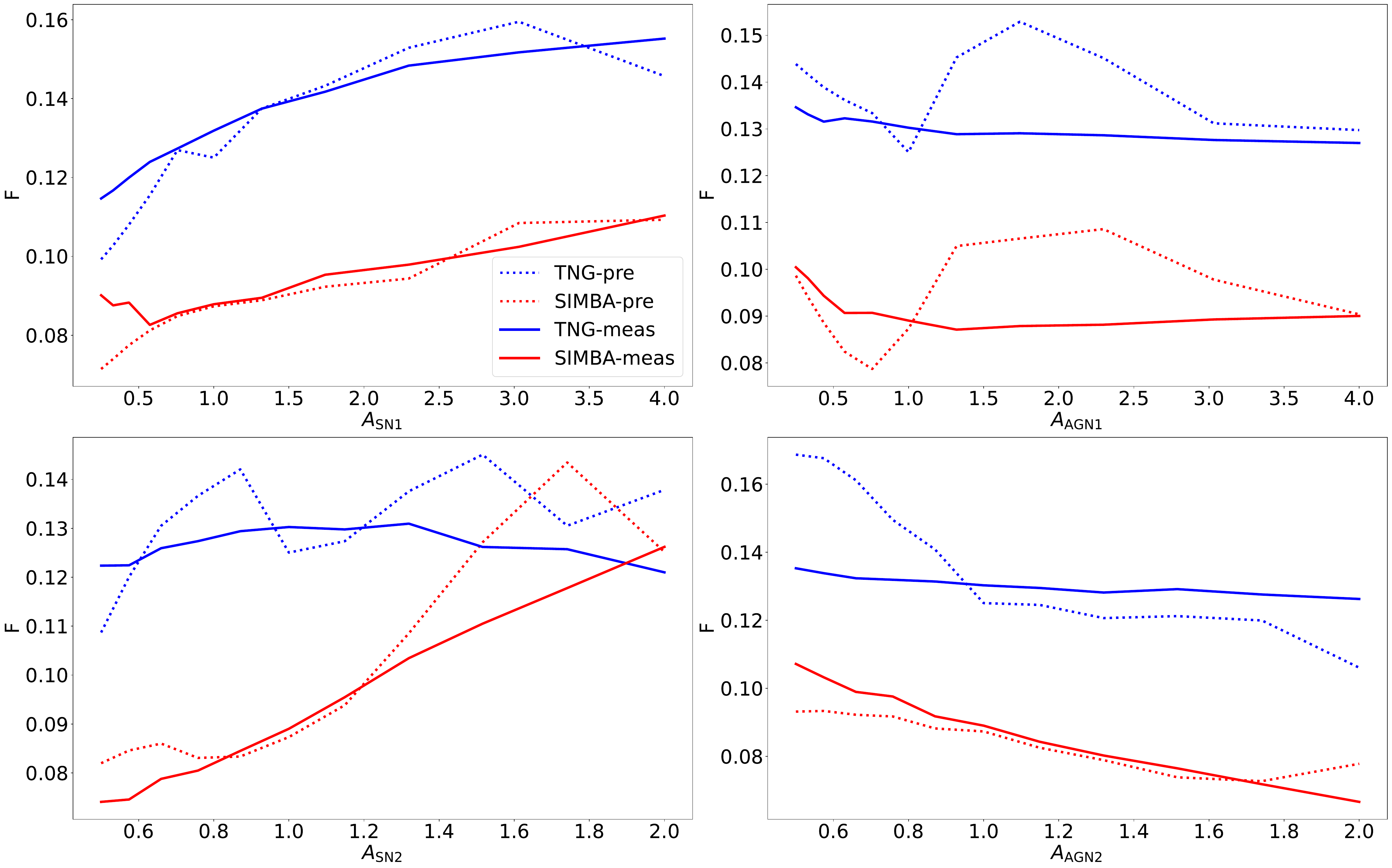}
    \caption{$F$ calculated from simulations in 1P set (solid lines) and predicted by our NN model (dotted lines) as a function of the feedback parameters in CAMELS for SIMBA (red) and TNG (blue). 
    NN has predicted the trends of these curves with varying astrophysical parameters.
    The results are different from those computed directly from 1P in other work. On the one hand, we are using different methods to compute DM from simulations. 
    On the other hand, we don't focus on the cosmic variance here, which could make some differences to the final predictions.}
    \label{fig_F_1P}
\end{figure*}
To quantify the strength of feedback, we study the $F$ parameter, which is sensitive to overall distribution of baryons, defined as
\begin{equation}\label{eq:F}
    F = \sigma_{\mathrm{DM}}(\Delta)\,z^{1/2},
\end{equation}
where the $z^{0.5}$ redshift scaling is due to the Poisson nature of random halo intersections.
In the FRB community, $F$ is often referred to as the `feedback parameter' controlling the scatter of \pdm, with the implicit assumption that stronger galaxy feedback increases the scatter of \pdm. 
As we shall see, this is an overly simplistic picture since the combination of different modes of feedback can affect $F$ (defined strictly as the scatter of \pdm{} through Eq.~\ref{eq:F}) in non-monotonic ways with respect to the true underlying feedback strengths.

Using our NN model, we explore the behavior of $F$ at redshift $z=0.5$ with varying astrophysical parameters while fixing $\Omega_m=0.3$ and $\sigma_8=0.8$. 
We denote these astrophysics-only parameters collectively as $ \theta_{\rm fb} $, where $A_{\rm SN1}, A_{\rm SN2}, A_{\rm AGN1}, A_{\rm AGN2} \in \theta_{\rm fb} $.
To explore $F$ across the $\theta_{\rm fb}$ parameter space, we generate $A_\mathrm{SN1}, A_\mathrm{AGN1}$ from 0.26 to 3.98 in steps of 0.02 and $A_\mathrm{SN2}, A_\mathrm{AGN2}$ from 0.51 to 1.97 in steps of 0.02, with all these parameters within the range of corresponding parameters in LH. 

We list in Table~\ref{tab:F} the values of $F$ when all the feedback parameters are set to their minimum or maximum within the allowed range in the CAMELS-SIMBA and CAMELS-TNG suites, respectively.
While one might have naively assumed that maximizing all the CAMELS feedback parameters would also maximize the resulting $F$, there are non-trivial interactions between the feedback processes parameters such that this is not the case. 
For instance, in SIMBA, stronger SNe feedback will suppress black hole growth and results in weaker effective impact of AGN feedback \citep{vanDaalen2011,Delgado2023}.
Table~\ref{tab:F} also shows the maximal and minimal $F$ at $z=0.5$, within our allowed range of $\theta_{\rm fb}$. 
In the case of CAMELS-SIMBA, the maximal $F=0.332$ is achieved by minimizing the parameters $A_{\rm SN1}$ and $A_{\rm AGN1}$ while maintaining a relatively strong $A_{\rm AGN2}$.
In CAMELS-TNG, on the other hand, the maximum feedback parameter of $F=0.309$ occurs when $A_{\rm SN1}$ is set to its maximum. 
The differences are partly due to the joint influence of feedback mechanism, while in SIMBA, SNe feedback has been found to suppress AGN feedback \citep{vanDaalen2011,Delgado2023}. 

We also noticed that for some feedback parameters, the behavior of $ F(\theta)$ from our NN model is different from the results reported by \citet{Medlock2024}. 
We show our corresponding results in Figure \ref{fig_F_1P}, which compares the predicted $F$ of the NN model with the $F$ computed from CAMELS-1P sets, varying each of the astrophysical parameters $[A_\mathrm{SN1},A_\mathrm{SN2},A_\mathrm{AGN1},A_\mathrm{AGN2}]$ at a time while fixing the others (including the two cosmological parameters) at the fiducial value. 
Note that the $F$ trends from the NN model are noisier than that computed directly from the 1P. This is because the 1P simulations were all computed with the same set of initial conditions whereas all the LH boxes have different initial conditions in addition to different model parameters $\theta$, so there is some additional sample variance in the latter due to different realizations of the underlying large-scale structure.
Especially in the SIMBA suites, we find that the increasing of $A_\mathrm{SN2}$, $F$ also increases, while $F$ decreases when $A_\mathrm{AGN1}$ increases, which is opposite from the trends found by \citet{Medlock2024}. 
In our analysis, the increase of SNe feedback leads to smaller $F$ (weaker feedback) as SNe feedback will suppress black hole growth and result in weaker effective impact of AGN feedback, which is also consistent with other results \citep{vanDaalen2011,Delgado2023}.
This is because we use different methods to compute $p(\mathrm{DM})$, which will lead to different estimates of $\sigma_{\mathrm{DM}}$.
\begin{table}\label{tab:F}
\caption{The values of $\mathbf{\theta}$ for the maximum and minimum values of $F(\mathbf{\theta})$ and the values of $F$ when all the feedback parameters are set to their minimum/maximum within the allowed range in CAMELS-SIMBA and CAMELS-TNG. Computed at a fiducial $z=0.5$}
\begin{tabular}{|l|l|l|l|l|l|l|}
\hline
                       & Description & $A_\mathrm{SN1}$ & $A_\mathrm{AGN1}$ & $A_\mathrm{SN2}$ & $A_\mathrm{AGN2}$ & $F$     \\ \hline
\multirow{4}{*}{SIMBA} & min($\thetafb$) & 0.26   & 0.26    & 0.51   & 0.51    & 0.143 \\ \cline{2-7} 
                       & max($\thetafb$) & 3.98   & 3.98    & 1.97   & 1.97    & 0.107 \\ \cline{2-7} 
                       & max($F$) & 0.26   & 0.26    & 1.71   & 1.01    & 0.332 \\ \cline{2-7} 
                       & min($F$) & 1.7    & 0.62    & 0.51   & 1.97    & 0.056 \\ \hline
\multirow{4}{*}{TNG}   & min($\thetafb$) & 0.26   & 0.26    & 0.51   & 0.51    & 0.168 \\ \cline{2-7} 
                       & max($\thetafb$) & 3.98   & 3.98    & 1.97   & 1.97    & 0.209 \\ \cline{2-7} 
                       & max($F$) & 3.98   & 0.64    & 0.53   & 0.51    & 0.309 \\ \cline{2-7} 
                       & min($F$) & 0.26   & 1.00    & 0.93   & 1.97    & 0.087 \\ \hline
\end{tabular}
\end{table}

Figure \ref{fig_DM_z} shows the Macquart (DM-$z$) relation for the CAMELS parameters that lead to minimum and maximum values of $F$. 
Once the cosmological parameters $\Omega_\mathrm{m}$ and $\sigma_8$ are fixed, the mean Macquart relation tends not to depend on astrophysical parameters --- this is to be expected since the Macquart relation depends primarily on $\Omega_b$ with only a slight modulation by the fraction of baryons that collapse into stars, galaxies and black holes.
This particularly leads to differences in the Macquart relation between the SIMBA and TNG models.
However, the feedback parameters significantly change the scatter of the DM-z relations within each of the simulation suites, with factors of $\sim 5$ difference in $F$ for SIMBA and factor of $\sim 3$ for TNG.
It is also clear that the inner 68th percentile of the distributions changes be only a factor of $\sim 1.5-2$ between the minimal and maximal $F$ models, with the differences in $F$ being driven most strongly by the outlying distributions as shown by the dashed lines that indicate the 0.15th and 99.85th percentiles.
We also notice that especially for SIMBA, the computed $\langle \mathrm{DM} \rangle$ in our work differs from that reported \citet{Medlock2024}, which might be due to different methods for computing DM.
For instance, at $z=1$, we find $\langle DM \rangle\sim800\,\mathrm{pc}\,\mathrm{cm}^{-3}$ using our method but \citet{Medlock2024} find $\langle DM \rangle\sim1002.5\,\mathrm{pc}\,\mathrm{cm}^{-3}$. 
Meanwhile, for TNG we used a different package to calculate DM, yielding $\langle DM \rangle\sim1036.96\,\mathrm{pc}\,\mathrm{cm}^{-3}$ but $\langle DM \rangle\sim1022.6\,\mathrm{pc}\,\mathrm{cm}^{-3}$ from \citet{Medlock2024} at $z=1$, which are comparatively closer and  also close to \citet{Walker2024} with $\langle DM \rangle\sim1020.60\,\mathrm{pc}\,\mathrm{cm}^{-3}$ at $z=1$ in TNG-300.
These percent-level differences are unsurprising, as they are from different box sizes and initial conditions (as well as resolution, in the case of \citealt{Walker2024}) albeit from the same TNG family.
However, in another study \citet{Zhang2021} found $\langle DM \rangle\sim892\,\mathrm{pc}\,\mathrm{cm}^{-3}$ at $z=1$ from on IllustrisTNG-100 model, which is smaller than the above TNG results.
Differences in $\langle DM\rangle$ of order $\sim 10-20\%$ can therefore arise even from similar simulations depending on the technique or code used to calculate DM.
A comparison project would be desirable to clarify these differences, which we will defer to future work.

We note that even the largest possible $F$ values from our model are small compared to existing observational constraints. 
For instance, \citet{Baptista2024} found $F = 0.331^{+0.271}_{-0.112}$ for their 76-FRB sample (assuming a uniform $H_0$ prior). 
Their most likely value is nearly the same as the maximum possible $F$ found in our SIMBA model (Table~\ref{tab:F}), and larger than our maximum TNG value.
We believe that the generally low values of $F$ in the CAMELS-based models can be attributed to the small simulation box size of $L=25\,\cMpc$.
This results in a very limited or non-existent representation of massive halos ($M_{\rm halo} \gtrsim 10^{14}\,M_\odot$) that reside primarily in overdensities, and conversely also cosmic voids in underdensities that can span tens of Megaparsecs (i.e. equivalent volumes to each CAMELS box). The dynamic range of gas and matter densities in CAMELS is thus significantly smaller than it would be in the real universe, leading to an artificially low scatter in the resulting \pdm.
While \citet{Medlock2024} used the \citet{Baptista2024} measurements to set some limits using their CAMELS 1P model, we believe it is premature to use CAMELS to make observational comparisons on the FRB \pdm{}.

\begin{figure}\label{fig_DM_z}
    \centering
    \includegraphics[width=0.46\textwidth]{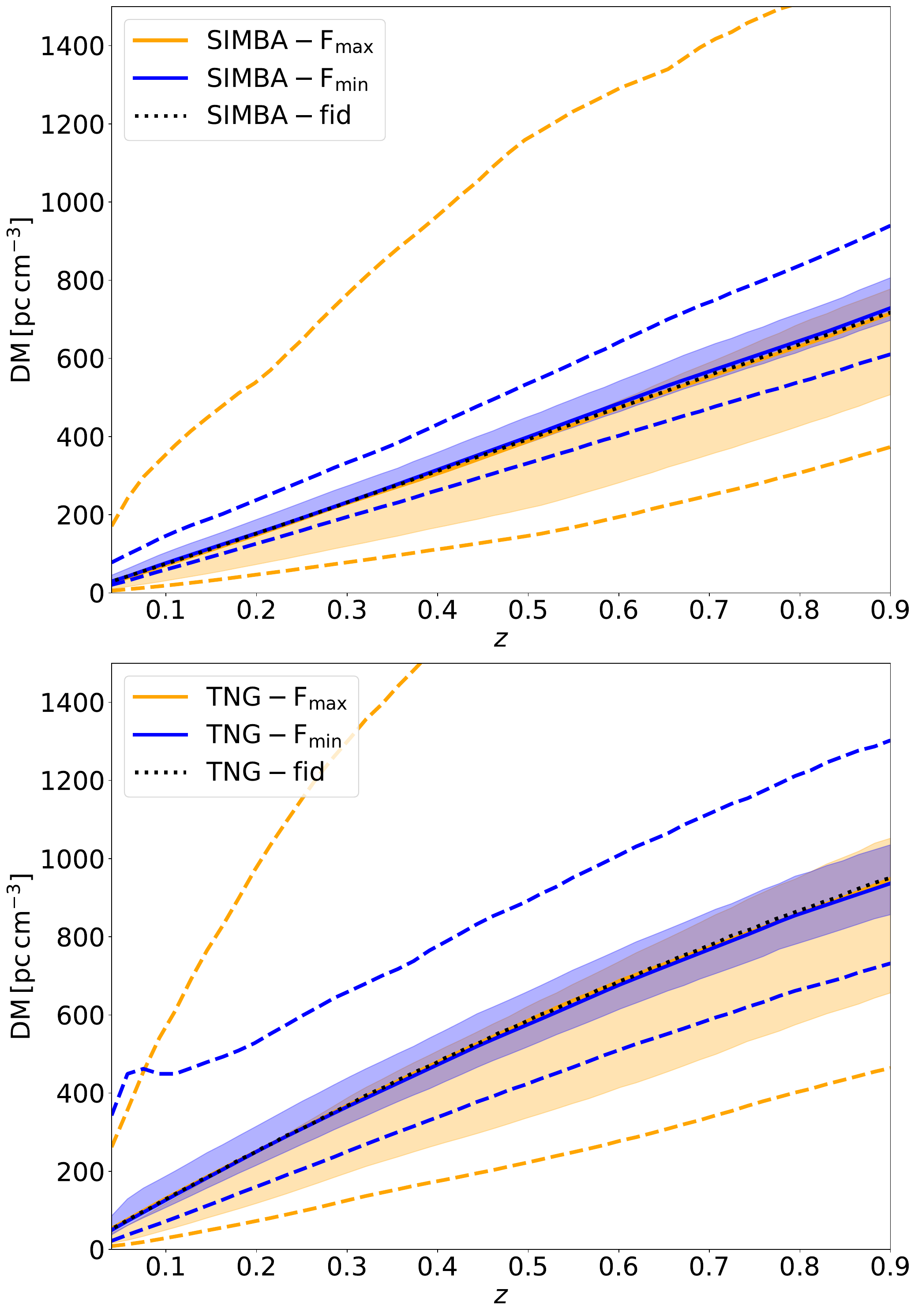}
    \caption{DM-$z$ relation with astrophysical parameters giving minimum (blue line) and maximum (red line) values of F for SIMBA (upper panel) and TNG (lower panel) model, with fixed $\Omega_m$ and $\sigma_8$. 
    The balck dotted line is the DM-$z$ relation with fiducial $\theta$.
    The solid lines are the $\langle DM \rangle$, while the shaded regions represent the 16th and 84th percentiles of the distributions. The dashed lines with the same colors indicate the 0.15th and 98.85th percentile outliers. 
    While the mean DM-z relations are quite similar within each suite, there are considerably differences in the scatter primarily driven by the outliers.
    }
    \label{fig_DM_z}
\end{figure}

\section{Discussion and Conclusion}\label{sec:conclusion}

In this work, we study the distribution of the FRB cosmic dispersion measure, $p(\mathrm{DM})$, and its dependence on feedback parameters in the CAMELS-SIMBA and CAMELS-TNG cosmological hydrodynamical simulation suites. 
The crucial element was to train a neural network model that could emulate the \pdm{} in CAMELS as a function of arbitrary redshift (at $z<1$) and underlying simulation parameters.
We list the methods and our main results below.

\begin{itemize}
    \item We compute the DMs in the simulation boxes using the pygad package for CAMELS-SIMBA and \texttt{temet} for CAMELS-TNG, respectively. 
    We note that different methods computing \pdm{} can give different results but we leave the discussion of this to future work.
    \item We apply a commonly-used functional form (Equation~\ref{equ_dm_dis}) to fit the $p(\mathrm{DM})$ computed from the simulations, which compresses the distribution to only 4 free parameters, i.e. $\left \langle \rm{DM} \right \rangle$, $\sigma_\mathrm{DM}$, $\alpha$ and $\beta$. (Figure \ref{fig_SIMBA_MCMC_fitting})
    \item We apply a neural network on \pdm, training the mapping between the CAMELS parameters $\theta$ to parameters describing $p(\mathrm{DM})$. 
    We find that NN makes good predictions for $p(\mathrm{DM})$ given cosmological and astrophysical parameters. (Figure \ref{fig_DM_results})
    \item We study the $F$ parameter using the model predictions for $\rm\sigma_{DM}$. 
    While the interplay between SNe feedback and AGN feedback is complicated and $F$ may not monotonically depend on every single feedback parameter, some intermediate points in astrophysical parameter space give minimum and maximum values of $F$, corresponding to stronger and weaker aggregate gas ejection from galactic haloes. (Figure \ref{fig_F_1P} and Table \ref{tab:F})
    \item The largest values of $F$ we find within the model is not large in comparison with observational constraints (e.g. \citealt{Baptista2024}), suggesting that the simulation box volumes in CAMELS is inadequate to capture the dynamic range of gas densities in the ICM, the CGM, and the IGM.
\end{itemize}    

Some previous efforts in the literature have already investigated the properties of baryons in CAMELS. 
For instance, \citet{Nicola2022} investigated the electron density auto-power spectrum as a probe of baryonic feedback. 
\citet{Gebhardt2024} investigated the redistribution of baryons owing to gravitational dynamics and feedback process, claiming that the increasing of AGN feedback efficiency increases the spread
of gas and stellar feedback has a contrary effect compared to AGN feedback. 
These studies, however, were focused on studying the behavior of the gas distributions within the simulations with only indirect references to observational probes.
More recently, \citet{Medlock2024} investigated the behavior of FRB DMs in CAMELS, but they only analyzed the limited 1P suite which all have the same initial conditions as well as only exploring the parameter space along one dimension at a time. 
Our goal in building the NN model of FRB DM from the full CAMELS LH suite was originally motivated to build a tool that could directly be used to interpret observational constraints such as those from \citet{Baptista2024}.
We do so using well-established NN packages in Python, and show that it performs well for both CAMELS-SIMBA and CAMELS-TNG.

However, there are significant limitations with our model, primarily inherited from the nature of the CAMELS simulations. 
For example, the size of the simulation box is too small to capture the cosmic variance, which can have significant effects on many measured quantities, though previous works in CAMELS have tried to find some predictors of cosmic variance or methods to mitigate them (e.g. \citet{Nicola2022}, \citet{Thiele2022}, \citet{Delgado2023}), but relevant extensions to our DM model is beyond the scope of this work.

While we conclude that models based on the existing CAMELS suite are not suitable for interpreting observational FRB DM results due to the small simulation box size, we nevertheless find that our NN model performs well for predicting the \pdm{} of CAMELS as a function of underlying parameters, both cosmological and astrophysical.
Future CAMELS-like simulation suites that cover larger simulation volumes ($L=50\,\cMpc$ or greater), or alternatively sampling a greater diversity of overdensity environments (e.g. \citet{Lee2024}, but additionally incorporating underdensities) will enable directly modeling of the observed Macquart relation.

Recently, the FLIMFLAM-like foreground mapping technique that directly models individual FRB foreground contributions \citep{Lee2022,Huang2024} has been shown to be a promising approach for studying the effect of feedback on the IGM and the CGM, specifically by constraining $f_{\rm igm}$ and $f_{\rm cgm}$, the cosmic baryon fractions residing in the IGM and the CGM respectively \citep{Khrykin2024a,Khrykin2024b}.
In a companion paper, we will present a NN model for $f_{\rm igm}$ and $f_{\rm cgm}$ derived from the CAMELS LH suite, using an analogous method to this paper.

However, the number of FRBs that do \textit{not} have foreground data will always exceed FLIMFLAM-like data samples, and direct modeling of \pdm{} will be required. 
This work points the way towards tying \pdm{} (or equivalently the Macquart relation) directly to feedback parameters in cosmological hydrodynamical simulations.


\section*{Acknowledgements}
We are grateful to Francisco Villaescusa-Navarro with assistance on the CAMELS suite, as well as Daniele Sorini and Dylan Nelson for their simulation tools. QG acknowledges support from USTC Fellowship for International Cooperation and the University of Tokyo Global Science Graduate Course program.
Kavli IPMU is supported by the World Premier International Research Center Initiative (WPI), MEXT, Japan.

\section*{Data Availability}
The code and models in this study is publicly available on GitHub \url{https://github.com/guoqigithub/DM_ML}.

\bibliographystyle{mnras}
\bibliography{example} 



\bsp	
\label{lastpage}
\end{document}